\documentclass[10pt,letterpaper]{article}
\usepackage[top=0.85in,left=2.75in,footskip=0.75in,marginparwidth=2in]{geometry}

\usepackage[utf8]{inputenc}

\usepackage{cite}

\usepackage{nameref,hyperref}

\usepackage[right]{lineno}

\usepackage{microtype}
\DisableLigatures[f]{encoding = *, family = * }

\raggedright
\usepackage{changepage}

\usepackage[aboveskip=1pt,labelfont=bf,labelsep=period,singlelinecheck=off]{caption}

\makeatletter
\renewcommand{\@biblabel}[1]{\quad#1.}
\makeatother

\usepackage{lastpage,fancyhdr,graphicx}
\usepackage{epstopdf}
\fancyheadoffset[L]{2.25in}
\fancyfootoffset[L]{2.25in}

\usepackage{color}

\definecolor{Gray}{gray}{.25}

\usepackage{graphicx}

\usepackage{sidecap}

\usepackage{wrapfig}
\usepackage[pscoord]{eso-pic}
\usepackage[fulladjust]{marginnote}
\reversemarginpar

\usepackage{amsmath}

\begin{document}
\vspace*{0.35in}

\begin{flushleft}
{\Large
\textbf\newline{Multilayer Networks in Neuroimaging}
}
\newline
\\
Vesna Vuksanovi\'c\textsuperscript{*},

\bigskip
\bf{} Health Data Science, Swansea University Medical School, Swansea University, UK
\bigskip
* vesna.vuksanovic@swansea.ac.uk

\end{flushleft}
\section*{Abstract} 
 Recent advances in network science, applied to \textit{in vivo} brain recordings, have paved the way for better understanding of the structure and function of the brain. However, despite its obvious usefulness in neuroscience, traditional network science lacks tools for --- so important --- simultaneous investigation of the inter-relationship between the two domains. In this chapter, I explore the increasing role of multilayer networks in building brain generative models and abilities of such models to uncover the full information about the brain complex spatiotemporal interactions that span across multiple scales and modalities. First, I begin with the theoretical foundation of brain networks accompanied by a brief overview of traditional networks and their role in constructing multilayer network models. Then, I delve into the applications of multilayer networks in neuroscience, particularly in deciphering structure-function relationship, modelling diseases, and integrating multi-scale and multi-modal data. Finally, I demonstrate how incorporating the multilayer framework into network neuroscience has brought to light previously hidden features of brain networks and, how multilayer networks can provide new insights and a description of the structure and function of the brain.

\section*{Introduction}
Network science provides a powerful tool for unraveling complex connectivity patterns in the brain [see, e.g, \cite{bassett2017network,stam2014modern} for review] and plays a crucial role in mapping its structural, and functional interactions [see, for example, \cite{avena2018communication}.] Until recently, most of the research on brain networks was based on traditional approaches of graph theory, where a typical network represents pair-wise relationships (links) between network units (nodes) \cite{rubinov2010complex}. While a such network representation has enabled years of progress in understanding connectivity patterns in the healthy and diseased human brain, recent work suggests that the description of the brain as a (monolayered) network may still be an oversimplification. Traditional representations of the brain as a single-layer network provide only limited representation of its structural and functional complexity.

Complex networks of interactions include a variety -- time-varying, dynamic -- connections across brain sub-systems (e.g. metabolic, cellular or regional) \cite{ritter2013virtual,patow2024whole}. The connectivity architecture of such connections exhibits an extraordinary level of “multiplexity” --- i.e. different type of interactions among its interacting units that span multiple scales --- spatial, temporal and topological. This has prompted a question whether the full information about brain networks can be better studied using the multilayer networks approach \cite{vaiana2020multilayer}. Although long present in many research fields to address diverse topics such as ecology \cite{pilosof2017multilayer}, transportation \cite{kurant2006layered}, animal behaviour and biology [see, e.g. \cite{finn2019use,de2016mapping}], or protein interactions \cite{ou2014detecting}, studies using multilayer methods for analysis of brain networks are still but a few \cite{de2017multilayer}. 

Here, I will discuss major advantages but also challenges in utilising the  multilayer methods in analysing  complex patterns of interactions in the brain. Particular attention will be paid to their potential for applications in exploring structure-function relationship, modelling brain diseases, and integrating multi-modal, multi-scale, big datasets in neuroimaging. The main focus will be on the so-called large-scale, or whole brain networks based on experimental recordings using Magnetic Resonance Imaging (MRI), including positron emission tomography (PET) and functional magnetic resonance imaging (fMRI), which allowed for the \textit{in-vivo} characterisation of the neural correlates underlying sensory,
cognitive, or behavioural tasks. \\

\begin{figure}
\includegraphics[width=0.99\textwidth]{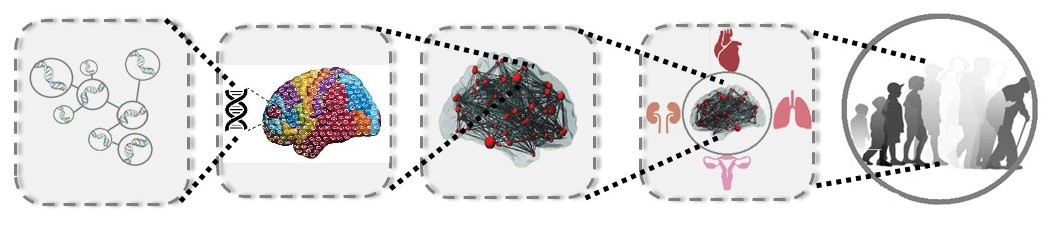}
\caption{Brain as a complex system at scales of organization and  interactions across different systems and modalities. 
\label{fig_1}}
\end{figure}

\section*{Multilayer networks: the theoretical background}
Intuitively, a multilayer network can be described a ‘network of networks’ or an 'interconnected network' -- a network comprised of
multiple interconnected layers, each characterising a different aspect of the same system. Different layers of connectivity arise naturally in natural, as well as in human-made systems. For example, relationships between humans include different types of interactions in the same network — such as relationships between family members, friends, and coworkers, which constitute different layers of a social system \cite{murase2014multilayer}. Other examples include transportation \cite{kurant2006layered}, ecology \cite{pilosof2017multilayer}, biology \cite{finn2019use}, and numerous other areas [see, e.g. \cite{boccaletti2014structure} for review]. \par 
The potential of multilayer networks to represent complex systems more accurately than it was previously possible has led to an increasing interest in physics of these networks \cite{de2013mathematical}. One significant application in physics emerges from the study of spreading processes \cite{de2016physics}, which have applications, e.g., in modelling traffic flows in transportation networks \cite{aleta2017multilayer} or the transmission of information and diseases in social networks \cite{wang2015coupled}. When spreading processes are coupled within a multilayered network approach to their analysis, this coupling can lead to some intriguing phenomena, which cannot be modelled within the single-layer network framework. In multilayer networks, the onset and dynamic of one process can depend on the onset of the other one, such as the emergence of critical points in phase diagrams, revealing distinct regimes where the criticality of the dynamics becomes intertwined or independent \cite{granell2013dynamical}. Furthermore, multilayer structures can foster cooperative behavior in structured populations \cite{li2020evolution}. By providing a platform for interactions across different layers, multilayer networks offer a novel mechanism for promoting cooperation even in complex environments \cite{duh2019assortativity}.

\subsection*{Monolayer Brain Networks}
One of the mathematical frameworks for studying the human brain structural (and functional) organization is graph theory. The brain network (graph) is modeled as a set of nodes and edges. Nodes and edges are elementary building blocks of networks and the definition of a node or an edge is of critical importance to the resulting brain network models \cite{zalesky2010,butts2009revisiting}. The arrangement of nodes and edges defines the organization of the network, whose topology is quantified using statistical tools of graph theory. Another major property of brain networks is the discovery that they are modular by their topological organization -- they can be decomposed into groups of nodes that are more densely connected to each other than with the rest of the network. A challenging question in the field of large-scale MRI-based brain network analysis is: How to define meaningful nodes for a brain network? The solutions range from defining nodes using the native resolution of the MRI technique (i.e., voxel-wise resolution) \cite{van2008small}, validated parcellations of the cortex based on anatomical or functional landmarks \cite{desikan2018release,tzourio2002} to using random parcellation to ensure equal size for each node \cite{fornito2010b,Hagmann2008}. More data-driven approaches include connectivity-defined nodes \cite{glasser2016multi}, multivariate decomposition of MRI signal (using statistical techniques such as independent component analysis) \cite{power2011} or a priori definition of nodes based on meta-analysis \cite{dosenbach2010prediction}. 

Most studies on functional and anatomical MRI networks use validated parcellations (brain atlases) to define nodes. The advantage of these methods is that they are informed by measures of brain function and anatomy and tailored to test specific hypothesis about brain networks of interest. The limitation is that they are not always transferable across different imaging modalities. However, the findings show consistency in the measures of network topology in different parcellation schemes and MRI modalities. In addition, the basic estimates of brain network organization, such as the degree of the node (number of nodal edges), the clustering (number of triangles in the network), or the path length (average number of edges between two nodes) are consistent across different parcellations with the same number of nodes. Modular organization, which is of interest here, is also consistent across anatomical and functional parcellations and modalities \cite{vuksanovic2019bcortical,chen2008revealing,meunier2009age}. Future work could corroborate these findings using available random cortex parcellations and multimodal MRI techniques. 

For example, in functional and morphological brain networks, edges are defined through an association matrix that captures relations (e.g. cross-correlation, mutual information etc.) between nodal features. The matrix maps all possible pair-wise statistical associations between either regional morphological features or time series of their activity. For the purpose of estimation of network topological organization, these matrices can be binarised -- mapping presence (and absence) of associations (edges); or weighted -- mapping strengths of association (edge strengths). There are differences in approaches to analyse these networks. Binarised networks are analysed over a range of binarisation thresholds to control for robustness and consistency of topological properties \cite{Bassett2009} and also for spurious/weak associations or noise \cite{vuk2019cortical}. Although arbitrary by its nature, threshold is usually determined by network's deviation from random, null-model topology \cite{rubinov2011} and the presence of small-world and scale-free topological properties \cite{Bassett2006}. Network edges can be weighted by the level (i.e., strength) of association between nodal interactions. In functional networks, edges are weighted by pair-wise temporal interactions, which are quantified either by correlation, coherence or synchronicity between time series \cite{power2011}. In anatomical networks, the edges are weighted by statistical associations (e.g., correlations) between different regional features: thickness, surface area, volume or curvature \cite{seidlitz2018morphometric,vuksanovic2019cortical,Bassett2008}, usually across groups of individuals. 

Although neither functional nor morphological correlation networks are constructed on direct neural (axonal) connections between the regions involved, both networks are largely constrained by the underlying structural network \cite{Honey2007}. For that reason, numerous studies have been focused on functional interactions that mirror the local (segregated) brain anatomy and axonal links between such interactions \cite{Honey2007}. However, the organizations of the two networks and their interactions (within the networks) suggest complex, multi-to-one function-structure mapping \cite{park2013structural,friston2010computational}. Using multilayer networks in studying the brain cytoarchitecture underpinnings of the complex patterns of structural-functional network associations, could produce a unified model of these associations. The relation between many functional corticocortical connections at scales could be modelled using multilayer networks. \par

Structural brain networks are usually derived from diffusion, dMRI data, by tracking the diffusion of water molecules along white matter tracts \cite{Hagmann2008}. dMRI exploits the anisotropic diffusion of water to reconstruct images that reveal the orientation and connectivity of white matter pathways. Similar to the brain functional networks, nodes typically correspond to distinct anatomical regions of the brain, defined based on standardised brain atlases, while links represent the white matter pathways that connect these regions \cite{}. Over the last decade, structural brain networks have been used to study normal brain development \cite{hagmann2012mr, huang2015development}, as well as for identifying alterations in the integrity of these networks associated with neurological disorders \cite{griffa2013structural}.  

\par

\subsection*{Structural representation of multilayer networks}
\label{Sec:structural}
This section focuses on a mathematical framework of the structural and organizational representation of multilayer networks that allow for the detection of network properties which are uncovered by traditional single-layer networks analyses. The definition of nodes and edges in multilayer networks differs from that of the traditional networks. For example, a node can be present in all layers of the network, but it does not have to be. Such organization allows for more creative approaches when defining network nodes and edges in the network. Within multilayer networks, three types of edges emerge: intra-layer edges linking nodes within the same layer, inter-layer edges between replica nodes across layers, and inter-layer edges connecting nodes representing distinct entities. Moreover, intra-layer and inter-layer edges encode relationships in fundamentally different ways, reflecting diverse function \cite{vaiana2020multilayer}. In a city transportation system, for instance, intra-layer edges denote connections between nodes of the same type (e.g., between different subway stations), while inter-layer edges link nodes of different types (e.g., between a subway station and an associated bus station) \cite{dickison2016multilayer}. In some instances, inter-layer and intra-layer edges may even be quantified using distinct physical units. For example, within a multilayer social network, an intra-layer edge could signify a friendship on Facebook, while an inter-layer edge might represent the probability of transitioning from Facebook to Twitter usage \cite{dickison2016multilayer}. In general, depending on the relative importance of intra-layer and inter-layer connections, a multilayer network can exhibit characteristics of either independent entities, with structurally decoupled layers, or a unified single-layer system where layers are practically indistinguishable. Some multilayer networks even exhibit a distinct transition between these two regimes. \par

Mathematically, the interactions in a multilayer network can be described by elements $m^{j\beta}_{i\alpha}$ of a 4th-order tensor $M$, known as the multilayer adjacency tensor. This tensor encodes the relationships between any node $i$ in layer $\alpha$ and any node $j$ in layer $\beta$ within the system, where $i, j \in \{1, 2, . . . , N\}$ and $\alpha$, $\beta \in \{1, 2, . . . , M\}$. Here, $N$ represents the number of nodes in the network, and $M$ denotes the number of layers. By describing the connectivity of nodes and layers using a tensor, novel measures can be defined to characterise the structural complexity of multilayer networks \cite{presigny2022colloquium}. In the general multilayer-network framework, each node can belong to any subset of the layers, and edges can encompass pairwise connections between all possible combinations of nodes and layers. (One can further generalize this framework to consider hyperedges that connect more than two nodes.) That is, a node u in layer $\alpha$ can be connected to any node v in any layer $\beta$. However, this process requires careful consideration, as simply extending concepts from monolayer networks can yield erroneous or nonsensical outcomes. The compact representation of multilayer networks provided by tensors enables greater abstraction, and has facilitated the advancement of mathematical frameworks for understanding complex systems \cite{kivela2014multilayer}. In the exploration of multilayer networks' structural properties, efforts have been dedicated to  simplifying the structural complexity of multilayer networks to reveal mesoscale structures, such as densely-connected communities of nodes, quantifying triadic relations like clustering, transitivity or motifs \cite{battiston2017multilayer}, and identifying key nodes based on various measures of importance \cite{yuvaraj2021topological, de2016mapping}. \par

An alternative approach, to the tensor representation, involves using sets of adjacency matrices to generalise concepts from monolayer to multilayer networks. This approach is based on the familiarity and the convenience of flattening adjacency tensors into matrices, also known as "supra-adjacency matrices," for computational purposes \cite{presigny2022colloquium}. The approach is used most frequently in neuroscience. In this case, the element of the supra-adjacency matrix $A$, $a_{ij}^{\alpha \beta} = a_{ji}^{\alpha \beta}$, describes interactions of node $i$ in layer $\alpha$ to node $j$ in layer $\beta$ of a given multilayer network. Network metrics, derived from the matrix $A$, have their equivalents in monolayer networks. For example, the equivalent of the \textit{node degree} is the so called overlapping degree or node strength, which sums the weighted degrees of node $i$ across all layers. Some other, popular in neuroscience, network metrics are: the \textit{multilayer participation index} \cite{battiston2017multilayer}, which quantifies nodal participation in interactions across layers; \textit{triangles}, i.e. triads of interconnected nodes based on the probability of forming triangles by means of links belonging to two different layers \cite{cozzo2015structure}. At the mesoscale level of network topology, one can calculate: \textit{motifs} [see, e.g. \cite{sporns2004motifs}]; \textit{modularity}, which captures network modules (communities) across all the layers simultaneously \cite{Mucha2010}; in temporal networks one can also extract \textit{flexibility}, which quantifies how often a node changes its modular allegiance acros layers \cite{bassett2011dynamic}. Several other metrics have also been used in neuroscience [see, e.g. \cite{presigny2022colloquium}], depending on the research question.  \par

The structural characteristics of multilayer networks are heavily influenced by how layers are interconnected to form (a cohesive) multilayer structure. Inter-layer edges are the links that bind layers together, encoding both structural and dynamical aspects of the system. Their presence or absence can leads to intriguing structural and dynamical phenomena. For example, in multimodal transportation systems, where layers represent different transportation modes, the weight of inter-layer connections may signify economic or temporal costs associated with switching between modes \cite{alessandretti2023multimodal}. In multilayer social networks, inter-layer connections models enable an interesting scenario in which many different transitions between the localisation in different layers are observed due to the multiple competition between them \cite{gomez2015layer}. More about the dynamic of multilayer networks in the next section~\ref{sec:dynammic}.

\subsection*{Dynamics of multilayer networks}
\label{sec:dynammic}
Similar to the existence of multiple topology, there are two types of dynamical processes in multilayer networks: (i) single dynamical processes unfolding on the top of interconnected structure of the multilayer network, and (ii) mixed or coupled dynamics, where multiple processes interact across layers through inter-layer connections. Most of the research on the dynamics of multilayer networks explore the role of inter-layer connections, in terms of interaction strength between layers and inter-layer communities. 
However, the behavior of single-layer processes is shaped by both intra-layer topology -- which is similar to traditional network analysis -- and inter-layer topology, representing interactions between nodes across layers \cite{de2016physics}. Such behaviour has lead to the discovery of many interesting phenomena related to network dynamic.  \par 

The study of diffusion, the simplest dynamical processes, in multilayer networks has uncovered an unexpected phenomenon: diffusion can manifest more rapidly in a multiplex network than in any of its constituent layers examined in isolation [see, e.g. \cite{de2016physics}]. Another example of discrete dynamical processes is random walk, commonly employed to model Markovian dynamics on monolayer networks and have offered valuable insights in the dynamic of spread in a network \cite{de2016physics}. In a random walk, a discretised form of diffusion, a walker traverses between nodes via available connections. However, in a multilayer network, these connections extend to layer-switching through inter-layer edges, a transition absent in monolayer networks. This enrichment of random-walk dynamics by layer switching highlights a key physical insight: "navigability." Navigability, defined as the mean fraction of nodes visited by a random walker within a finite time, reveals an intriguing aspect of the interplay between multilayer structure and dynamic processes. Similar to continuous diffusion, navigability in multilayer networks can surpass that of an aggregated network of layers. This phenomenon indicates that multilayer networks exhibit greater resilience to uniformly random failures compared to their individual layers \cite{sole2016congestion}. This resilience emerges directly from the  interplay between the multilayer structure and the underlying dynamical processes. The phenomena like navigability, can shed light on the network's resilience and efficiency in the context of various perturbations and attacks , which have already been explored in monolayer networks for modelling disease spread \cite{boguna2009navigability,de2014navigability}. Another significant phenomenon to consider is congestion, which represents the delicate balance between flow across network structures and the capacity of those structures to accommodate such flow. While congestion in networks has been a subject of analysis in physics literature for many years \cite{arenas2003search}, its study in the context of multilayer networks is a relatively recent development \cite{sole2016congestion,manfredi2018mobility}. \par

The coupled dynamics of multilayer networks considers two fundamental effects: firstly, mutual excitation, where one process facilitates the spread of the other (e.g., one disease amplifying the infection of another); secondly, mutual inhibition, where one process hinders the spread of the other (e.g., disease A inhibiting the infection of disease B, or the dissemination of awareness about a disease curbing its spread) \cite{wang2015coupled,raj2018models}. The interaction between mutual excitation and inhibition yields a complex dynamic, which make such interplay an ideal candidate for modelling brain dynamic.  \par

\section*{Multilayer networks in neuroimaging} 
\label{Sec:dynamics}

Brain (structural and functional) interactions span across different levels of organization and scales. Each of such levels can be described by distinct layers of connectivity and activity (function), representing an ideal foundation for multiyear network modelling. For example, different network layers can represent functional networks, that is co-activation of multiple brain regions involved in specific cognitive tasks or affected by diseases \cite{}. Similarly, structural multilayers can map the physical connections, i.e., at scales connectivity of brain units. Layers can also describe multiple features that interact at different levels of organization \cite{}, thus bridging between different levels of information and properties. An example of this type of modelling are computational models that integrate information at the neuronal level to predict the whole brain dynamics \cite{Jirsa2010,deco2016metastability,Breakspear2010}.   

\begin{figure}
\includegraphics[width=0.99\textwidth]{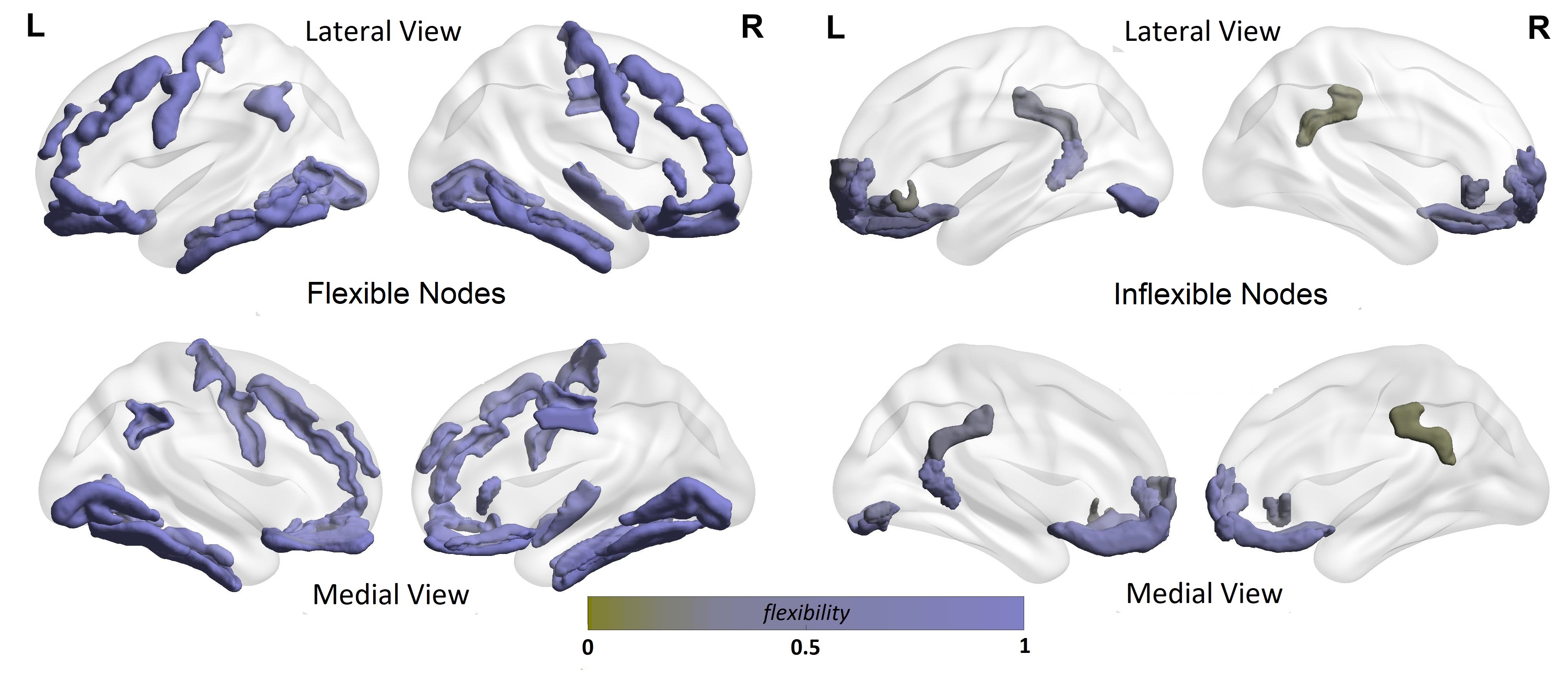}
\caption{Brain 'flexible' hubs in ageing revealed by multilayer network analysis. Modified from \cite{vuksanovic2022brain}}
\label{Fig:fig_2}
\end{figure}

Another, important, feature of multilayer networks in neuroimaging is in retaining information about individual brain networks. Until recently, the brain networks from neuroimaging data have usually been aggregated across individuals into a single-layer network, e.g. by averaging. With the advances of multilayer networks and network statistics designed to operate across layers, the full multi-dimensional information can be retained, allowing for a richer analysis that could not be obtained from the aggregated single-layer network. For example, in \cite{de2016mapping}, the authors show that a multilayer representation of a human brain network gives higher classification accuracy between healthy and schizophrenic patients than the single-layer and aggregate counterpart networks. Similarly, \cite{vuksanovic2022brain}, shows that a multilayer representation of a human brain networks gives a basis for understanding changes in its organization with ageing (see Fig.~\ref{Fig:fig_2}) or variations in IQ. 

The first topic in neuroimaging, which have used multilayer approach even before a formal mathematical framework was developed, was to consider time-varying functional networks from fMRI data \cite{braun2015dynamic}. In this case, the resulting multilayer network was a collection of adjacency matrices, each of which corresponds to a snapshot of brain activity over time -- each represents Mapping the changes in topology of these networks over time, have shown their biological relevance \cite{bassett2011dynamic}. Similarly, in EEG and MEG research communities, there is an increasing number of studies that use, so called, frequency-based decomposition of multilayer networks. This method is based on building the multilayer functional brain using the EEG/MEG signal from different brain regions, decomposed in the frequency domain. To build a multilayer network, each layer is represented by functional connectivity in different frequency bands [see, e.g. \cite{brookes2016multi,tewarie2016integrating}]. A more recent study, multilayer network approach has been shown to provide high-classification accuracy and sensitivity of subjects with cognitive impairments and AD \cite{echegoyen2021single}. 

Moreover, by integrating data from different imaging modalities, such as functional MRI, diffusion MRI, and magneto/electroencephalography (MEG/EEG), multilayer networks approach can be utilised to construct generative models that capture the dynamic interplay between brain units across multiple \textit{layers} of brain complexity  \cite{mandke2018comparing,puxeddu2020modular,puxeddu2021comprehensive}. The advantage of network theory is that is independent on (imaging) modality, i.e. the same mathematical framework can be applied independently on the modality that the data come from.  This modality-invariant characteristic opens numerous possibilities for studying interactions between different brain systems, and offers a distinct perspective on brain connectivity from single-network approach based on networks analysed using data from a single modality. However, despite the obvious promise of these approaches, a key issue in the application of network theory is comparison between groups or conditions. Similarly to mono-layer network approach, it is difficult to distinguish between alterations in the underlying brain connectivity caused by experimental manipulation or a disease process and those resulting from computational choices such as network thresholding  \cite{van2010comparing}. The choices can potentially influence false positives or false negatives when comparing brain networks across individuals and/or groups. Multilayer brain networks are not exempt from these biases and need a careful consideration when comparing their organization and metrics \cite{brookes2016multi,de2016physics, guillon2017loss}. \par
Neuroimaging models of brain dynamic and their multilayer networks representation, can provide better understanding of functional connectivity of the brain. In the context of functional connectivity (network) it is important to address the concept of correlations. In single-layer networks, it has been observed that there are correlations in the properties of connected nodes. For example, the degree of a node can exhibit either positive or negative correlations with the degree of its neighboring nodes. Positive correlations suggest that hubs tend to connect preferentially with other hubs, while negative correlations indicate a preference for hubs to connect with low-degree nodes. However, in multiplex networks, the notion of correlations becomes much richer compared to single-layer networks \cite{nicosia2015measuring}. While it is still possible to examine standard degree-degree correlations within each layer of the network, it becomes more intriguing to define correlations in a truly multiplex context. For instance, to investigate how a certain property of a node in one layer correlates with the same or other properties of the same node in another layer. A study by \cite{nicosia2015measuring}
provides a comprehensive examination of correlations of node properties in multiplex networks within the framework of neuroimaging models of brain dynamics. To achieve this, the authors follow a systematic approach commonly employed in complex network analysis:
i) empirical exploration of correlations in real multiplex brain networks derived from structural and functional MRI data, 
ii) introduction of various measures to characterise and quantify these correlations, offering insights into the organization of the brain's multilayer network and (iii) a series of models were proposed to replicate the observed correlations in real brain networks or assess their significance, thus contributing to our understanding of the underlying principles governing brain dynamics.

\subsection*{Structure-function relationship in the brain}

Functional networks, which represent patterns of neural activity distributed across the brain, are a long-term focus in neuroscience. The dynamical properties of these networks, especially their deviation from the comparatively static structural networks, are thought to be critical for higher cognitive functions. The interplay between structural networks and emergent functional activity, combined with the application of multimodal neuroimaging techniques and standard frequency band analyses, underscores the need for a multilayer network approach. \par 
Despite the innate relationship between structural and functional brain networks, their direct comparisons and simultaneous analysis poses challenges \cite{deco2016metastability,Vuksanovic2014}. While theoretical studies have made the progress in understanding the relationship between brain structure and dynamic \cite{Deco2011}, it is less clear how structural network facilitates the emergence of functional network. The correspondence between functional and structural brain networks remains an active area of research \cite{suarez2020linking}. Given that  functional networks are constrained by invariant structural connections, functional integration during various cognitive tasks necessarily imply many-to-one function-structure mapping \cite{friston2010computational}. In this context, modelling brain dynamics and organizational representation as a multilayer network from structural and functional brain networks constructed through MRI data offers a comprehensive approach to their \textit{simultaneous} analysis.

The non-trivial relationship between structural and functional connectivity poses many challenges about brain development, cognition, diseases, learning and aging \cite{puxeddu2020modular,vuksanovic2021brain}. 
A growing body of research addresses these challenges trough the prediction of functional connectivity from structural characteristics [see some early works based on the single layer approach, e.g. \cite{Honey2007,Hagmann2008,rubinov2011}]. The methods used for the prediction are based on the application of statistical tools \cite{vazquez2019gradients}, information flow framework for neural communication \cite{Deco2009} or considering the coupling between structure and function through non-linear dynamics \cite{Deco2009,Breakspear2010,ritter2013virtual}, or examining the effects of brain lesions on functional organization and behavior \cite{ghajari2017computational}. \par 
Beyond predicting one from the other, structural and functional networks can be jointly analysed to investigate shared organizational properties, such as modular structure, and their interrelationships. The tensor representation of multilayer networks provides a robust mathematical framework, enabling the extension of traditional complex network analysis techniques. Within the framework, the methods, such as detecting modular super-units and identifying central nodes, are instrumental in understanding the organizational structure of the human brain \cite{sporns2013structure,sporns2016modular,vuksanovic2022brain}. Modular topology is one ubiquitous characteristic of complex networks (including the human brain). Networks can be divided into modules by grouping the densely intra-connected sub-sets of nodes into a single sub-group (i.e., module). Algorithms for the division of a (real-world) network into modules are usually optimized to allow for sparse connections between groups (i.e., detection of overlapping communities) \cite{newman2006modularity,fortunato2016community}. Furthermore, detecting modules in the network may help to identify those nodes and their connections that may perform different functions with some degree of independence. At the same time, detecting modular structures that underpin specific function can be identified by characterizing interactions between those nodes that show relatively similar activity/dynamics \cite{fortunato2016community}. Likewise, meta-analysis on more than 1000 fMRI has shown the existence of  functional modules specialized for specific cognitive processes \cite{crossley2013cognitive}.

The brain appears to be divided into 'functional modules' whose intra-modular connectivity reflects the underlying structural (axonal) connections \cite{Honey2007}. However, although functional modules usually mirror local brain anatomy, they also incorporate long-range interactions (i.e., those between spatially distant brain areas) \cite{Fries2005,Vuksanovic2014,vuksanovic2015dynamic}. The modular topology of brain functional (MRI) networks is documented across different parcellations of the cortex (i.e., brain atlases) \cite{meunier2009age,bertolero2015modular,vuksanovic2019bcortical}. Modularity as a property of morphology has been widely studied in the context of evolution and development \cite{melo2016modularity}. Recent neuroimaging studies suggest modular organization of cortical morphology across regional thickness \cite{vuksanovic2019cortical,vuksanovic2019bcortical}, surface area \cite{sanabria2010surface} or volume \cite{Bassett2008}. There is consistency in the organization of these networks whether they are based on correlating these features across individuals within one group \cite{vuksanovic2019bcortical,sanabria2010surface,vavsa2017adolescent} or correlating regional features of an individual brain \cite{seidlitz2018morphometric}. The brain modular, yet integrated, functional organization lowers the wiring cost (i.e., the average length and number of connections) of the network \cite{Bassett2009a}, thus potentially lowering metabolic costs \cite{betzel2017modular} while providing more efficient information processing \cite{sporns2016modular}. More importantly, modularity, as mapped by large-scale brain fMRI networks, is cognitively and behaviorally relevant; for example, it correlates with variations in working memory \cite{yamashita2015predicting}. This property is evident in both anatomical and functional networks across different spatial and temporal scales \cite{bassett2017network}. 
\par
Brain networks can be divided into functionally relevant modules either into smaller communities composed of functionally specialized areas or large modules hypothesised to support complex cognitive functions. Many of tools used in these approaches are focused on analysing how information propagates through the multilayer system, making them well-suited for the structural analysis of brain dynamics \cite{de2014navigability,sole2016congestion}. However, while classical network concepts have been successfully extended to multilayer systems, approaches used to model the human brain predominantly rely on multiplex and interconnected multiplex topologies \cite{de2016mapping}. In these models, nodes are often replicated across multiple layers, each layer encoding different aspects of brain function. This organizational representation, derived from structural and functional brain networks constructed from MRI data, captures the complex interplay between brain regions and the dynamic flow of information within the multilayer network of the brain. \par

\section*{Integrating multi-modal, multi-scale data: a multilayer network approach}
In the past decades, several experimental measurements, based on structural and functional MRIs, diffusion tensor imaging (DTI), PET imaging, or electroencephalography (EEG) and magnetoencephalography (MEG), have been used for non-invasive reconstructions and recordings of brain structures and activity. The existence of such large and multi-modal data sets, providing snapshots of the brain at scales, have prompted appearance of quantitative models that involve explorations of interactions between multiple modalities in health [see, e.g. \cite{bassett2017network}] and disease [see, e.g., \cite{franzmeie2019functional,khan2020whole}]. 
The advantage of a such approach is to allow for multiple elements or interactions of a networked system to be analysed under a single model. Before I address multilayer networks from multimodal neuroimaging data, it should be noted that, multi-layer network models have been used for over a decade now to study the covariance structure of functional brain data, by analysing time-varying functional networks from fMRI data [see, e.g. \cite{bassett2011dynamic, Mucha2010})]. A similar approach has also been used to establish nodal modular allegiances across networks representing individual brains using structural and anatomical MRIs \cite{puxeddu2022multi,vuksanovic2022brain}.

Recently, a smaller number of studies have begun investigating multi-layer network models of the brain across multiple imaging modalities \cite{de2017multilayer, vaiana2020multilayer,puxeddu2022multi}. A first such attempt has been made using fMRI and DTI data, by constructing layers from the fMRI and DTI networks of a single subject
\cite{battiston2017multilayer,crofts2016structure,crofts2022structure}. In such models, the fMRI layer represents the fMRI network, where the nodes are brain regions, and edges are statistical associations between regional activity. The DTI layer represents the weights of connections of the white matter tracts between (the same) brain regions. In line with previous studies, the results confirm the complex relationships between structural connectivity and coupling of brain dynamics. Structural multi-layer communities differ from those usually obtained from single structural connectivity networks, suggesting that the multi-layer framework could be a more appropriate choice for the analysis of multimodal brain networks. \par

A comprehensive understanding of the brain require an integrated approach across various scales and levels of organization. This involves, for example, the interplay of genes and synapses, as well as the relationship between the structure and dynamics of the entire brain, which ultimately shapes diverse behavior and cognitive function \cite{presigny2022colloquium}. Multiscale brain modeling poses significant challenges, partly due to the difficulty of simultaneously accessing information across multiple scales and levels \cite{goubran2019multimodal,valk2017structural}. While there has been some progress in understanding the role of specific microcircuits in generating macroscale brain activity through biologically inspired computational models [see, e.g. \cite{Deco2011,Jirsa2010,Breakspear2010}], a thorough characterisation of how changes at one scale or level impact others remains elusive. The theoretical advancements of multilayer networks offer a robust framework for analysing and modelling interconnected systems with interactions within and between different layers of information. Recent progress includes characterising multiscale brain organization in terms of structure-function relationships \cite{schirner2018inferring}, oscillation frequencies \cite{deco2016metastability}, and temporal dynamics \cite{betzel2017multi}. In addition, the multilayer network properties have been used to identify multimodal network-based biomarkers of brain pathologies in Alzheimer’s disease \cite{guillon2017loss}. \par
I have presented some of conceptual frameworks, tools, and results that provide insights into how brain systems can be represented via the intra- and inter-layer network properties. However, research in this field is active, and many issues remain to be addressed to ultimately characterise the multiscale, multilevel brain organization \cite{vaiana2020multilayer, de2017multilayer, puxeddu2022multi presigny2022colloquium,vuksanovic2022brain}. Finally, each imaging modality offers a unique perspective on brain function or structure, and data fusion leverages the strengths of each modality and their interrelationships in a joint analysis. However, most studies favor only one data type or fail to combine modalities in an integrated manner, thereby missing significant changes that are only partially detected by individual modalities. Conversely, multimodal fusion provides a more comprehensive depiction of altered brain patterns and connectivity. This approach has demonstrated increasing utility in addressing both scientifically intriguing and clinically relevant questions, offering a more holistic understanding of brain dynamics.

\section*{Application of multilayer networks in modelling neurodegenerative disorders}
The application of monolayer network approaches has been a promising strategy for unraveling the complex role of molecular, cellular, and circuit-level changes associated with brain disorders [see, e.g. \cite{stam2014modern,fan2023macroscale}]. In this section I will explore the potential of multilayer network analysis in elucidating the pathophysiology, progression, and potential therapeutic targets of neurodegenerative, neurological, and psychiatric disorders. As I have already emphasised, research investigating these questions (i) spans multiple spatial and temporal scales, (ii) results in the use of multiple modalities to experimental data, and (iii) involves comparison across multiple subjects or between subject groups. A multilayer network is a promising tool for reconciling these different aspects in a coherent and consistent manner. Its main advantage is that the networks' nodes (e.g., brain regions) can be coherently linked across many modes of analysis resulting in a single structure which contains the full information of the data. There are two main pathways for the multilayer network approach to improve our understanding of neurodegeneration --- cellular heterogeneity and connectomics. \par

Common, underlying, characteristic of brain disorders, such as mental disorders \cite{richetto2021epigenetic}, epilepsy \cite{guerrini2003genetic} or dementia-causing disorders \cite{dujardin2020tau,rademakers2012advances} is molecular heterogeneity. For example, neurodegenerative disorders that cause dementia are, without exceptions, characterised by molecular heterogeneity. Such heterogeneity involves, e.g. aberrant protein aggregation \cite{dujardin2020tau}, synaptic dysfunction \cite{dujardin2020tau}, or neuronal loss \cite{west1994differences}. While traditional single-layer network models often overlook the multifaceted nature of neurodegeneration, by focusing on a single imaging modality; the multilayer networks can provide an integrated framework for diverse molecular and cellular sources of neurodegeneration. By representing different molecular layers (i.e., sources of heterogeneity) as interconnected networks, the multilayer models can capture the dynamic interactions across the scales that are involved in neurodegeneration. For example integrating gene expression data with multimodal imaging data \cite{thompson2014enigma, rubido2024genetic}, may lead to uncovering disease-associated connectome failures, protein dysregulation, and/or protein aggregation pathways \cite{raj2018models}.  \par

In addition, there is increasing evidence from multimodal studies that indicates that individuals with neurodegenerative disorders exhibit distinct morphological changes \cite{lieberman2001longitudinal,wurthmann1995brain}, connectivity patterns \cite{franzmeie2019functional,vuksanovic2019bcortical}, and functional alterations \cite{franzmeie2019functional,wang2020neuroimaging}. These changes are not easily discernible through separate unimodal analyses, which are typically performed in the majority of neuroimaging experiments. Thus, the application of multilayer network analyses to these multimodal characteristics could facilitate the identification of biomarkers for brain disorders diseases, expediting differential diagnosis and leading to more appropriate treatments and improved outcomes for patients. For example, mild cognitive impairment (MCI) is challenging to diagnose due to its subtle and often insignificant symptoms of cognitive impairment. Some studies have highlighted the potential of combining structural and functional data with multimodal classification techniques to enhance the accuracy and early detection of brain abnormalities in MCI \cite{fan2008spatial}.

Another aspect of the multilayer network approach that could be utilised in brain disorders involves focusing on the spatial and temporal dynamics of these networks in the context of disease [see, e.g. \cite{stam2014modern}]. In general, disease can impact brain networks through two mechanisms. Firstly, disease can directly affect the nodes of the network, leading to either localised or widespread functional changes. This impairment can propagate along neural connections to other brain regions \cite{franzmeie2019functional,franzmeier2020functional}, thereby amplifying the dysfunction. Secondly, certain diseases mainly target neural connections themselves, such as demyelination or axonal injury \cite{ettle2016oligodendroglia}. This disruption in connectivity results in anomalous information processing and widespread functional impairments, reflecting the underlying aberrant network dynamics. \cite{raj2018models,raj2012network,rubido2024genetic}. For example, utilising a multilayer network approach to identify the type of disease dynamics can reveal  pathological signatures across different neurodegenerative diseases, revealing convergent dysregulation of brain subsystems involved in these processes. However, clustering of multilayer networks, especially using  information on higher-order interactions of system entities, still
remains in its infancy. Partially because higher-order network metrics are often the key in such multilayer network applications \cite{yuvaraj2021topological}. Developing
optimal partitioning of critical topological features can help to isolate unhealthy components of the networks targeted by disease and identification of multiple brain regions affected
by trauma or mental disorders.

\section*{Conclusion and Future Directions}

In this chapter, I have presented novel conceptual insights, tools, and results that offer new perspectives to model brain systems and interactions between and within, using the multilayer network approach. Multilayer networks represent a simple yet flexible data structure that enables unique quantitative analysis of complex data, whilst preserving information and architecture otherwise lost with traditional statistical, network approaches. Retention of additional, hidden information, coupled with the development of novel network statistics, has successfully provided insights into the structure and function of the human brain that previously remained hidden. Furthermore, multilayer network approaches offer a comprehensive framework for unraveling the complexity of mechanisms and processes in neurodegenerative disorders, which can facilitate biomarker discovery and improve therapeutic strategies. By integrating diverse data sources and capturing spatial and temporal dynamics, multilayer network analysis holds significant promise for advancing our understanding of neurodegeneration and ultimately improving patient outcomes. Continued interdisciplinary collaborations and technological advancements will be essential to translate multilayer network findings into clinical practice. The following years will be crucial for elucidating how multilevel brain complexity and multilayer network tools can be synergistically merged to establish a new generation of network-based multiscale and multimodal models of brain organization. \par
Research in multilayer networks in neuroimaging is ongoing, and numerous questions remain to be addressed to fully characterise the multiscale, multilevel organization of the brain. For example, despite the promise of multilayer network analysis, several challenges remain in applying these approaches to neuroimaging data. Integration of heterogeneous data sources, standardisation of analytical methods, and validation of findings across independent cohorts are critical features of the process. Furthermore, the development of scalable computational tools and platforms for multilayer network analysis is needed to facilitate widespread adoption in the research community. I conclude this chapter by highlighting some key directions in the the advancement of multilayer network in neuroimaging particularly pertinent for addressing these challenges. \par


\nolinenumbers



\end{document}